\documentclass[11pt, graphicx]{article}
\usepackage{delarray}
\usepackage{epsfig}

\begin{document} 

\begin{center}
\Large\textbf {Particle Spectra Of Ultra-Relativistic Shock Waves\footnote{Some of these results were presented in the \textit {17$^{th}$ European Cosmic Ray Symposium}, 
July 2000, Lodz Poland}}
\end{center}

\begin{center}
\renewcommand{\thefootnote}{\fnsymbol{footnote}} 
{\Large{\textit{Athina Meli and John Quenby}}}
\end{center}
\begin{center}
{\large{Astrophysics Group, Blackett Laboratory}}
{\large{Imperial College of Science,Technology and Medicine, London, 
UK}}
\end{center}

\begin{abstract}
\small\textbf{ Numerical Monte Carlo simulations of the diffusive 
shock acceleration in the test particle limit are investigated. We simulate high relativistic 
flow astrophysical plasmas for upstream  $\gamma$ $\sim5$ and up to 
$\gamma$ $\sim1000$. These gamma values are relevant to the diffusive shock acceleration models of Active Galactic Nuclei (AGN) and Gamma Ray Bursts (GRBs). The spectral shape for the above high relativistic processes will be calculated. 
This work presents distinctive spectral shapes with plateau-like characteristics due to the 
relativistic effects and the spectral indexes are found to be as flat as the flow becomes 
relativistic; the latter is in a very good agreement with previous theoretical and numerical 
works concerning 
relativistic plasma flows.}
\end{abstract}

\begin{center}
{\large{\textit{\textbf {Introduction}}}}
\end{center}

{\large It is known that multiwavelength observations from both Gamma Ray Bursts 
(GRB) (Meszaros \& Rees, 1993) and a certain number of Active Galactic Nuclei (AGN), 
indicate as well as invoke, the presence of 
relativistic and ultrarelativistic shock waves. In these shocks the mechanism of 
diffusive 
shock acceleration at collisionless plasma shocks takes place (Peacock 1981) 
and explains the appearance of the non-thermal particle populations from these sources.
Specifically, authors such as Vietri (1995) and Waxman (1995) have pointed out that these 
extremely energetic events with 
hyper-relativistic flows ($\Gamma$ $\sim100-1000$) in GRBs, could efficiently accelerate 
protons which could be the source of the Ultra High Energy Cosmic Rays (UHECR).
It is an important fact that the analytical 
models of Vietri and results on UHECR, depend critically on  past (Lieu \& Quenby, 1989) pioneering 
large angle scattering numerical and analytical simulations of particle diffusion 
shock acceleration 
for $\Gamma$ flows $\sim3$ where they reported a spectral 
flattening when the flow becomes relativistic (Kirk \& Schneider, 1987 first noticed a spectral flattening). For $U_{1}$=0.96c, $U_{2}$=0.32c (that is compression ratio = 3) they find the spectral index to be flattened to around 1.2 in excellent agreement with theoretical and numerical
work of Kirk and Schneider (1987a,b).
The numerical results obtained by our current Monte Carlo simulations in the regime of 
diffusive shock
 acceleration for 
$\Gamma$$\sim 5-10^{3}$, are consistent again with the most plausible predictions regarding the
flatteness of the spectrum.
The first order Fermi acceleration theory for parallel non-relativistic flows,
yields a spectral index $\sim$2 for the differential spectrum dn/dE$\propto$$E^{-\alpha}$, where $\alpha=r+2/r+1$ and $r=u_{1}/u_{2}$. 
In the relativistic and ultra-relativistic limits for diffusive acceleration
many studies investigated the issue of the spectrum index's behaviour in relation to flow velocities and magnetic field irreqularities present in the relativistic plasma flow and do find a considerable spectral flattening as
the plasma flow becomes relativistic; \textit{but} the spectral shape at the source is not yet 
clearly understood. The aim of this work is to address the problem of the spectral shape at 
source for both large angle scattering  and pitch angle diffusion models in order to 
investigate the variations between the two. Monte Carlo numerical simulations have been 
used in the test-particle diffusive limit employing highly rleativistic flows. We compare our 
results with similar studies for $\gamma$$\sim$5 (Baring, 1999)
and find a good agreement concerning the spectral index and the spectral shape, for both large 
angle scattering and pitch angle diffusion mechanisms}.

\begin{center}
{\large{\textit{\textbf {Monte Carlo Method}}}}
\end{center}

{\large In our Monte Carlo code, the particle \textit{splitting} mechanism is used in order to
improve the statistics of the number of the remaining particles in the acceleration process. 
The injection of energetic seed particles ($\sim$$10^{7}$), with weight equal to 1.0, is far 
upstream. Particles are allowed  to scatter randomly in the respective fluid frames 
towards the shock. A guiding centre approximation is used  and the  
particle propagation  is followed in a 1D geometry. Each particle trajectory is been 
followed across the shock according to  known 'jump conditions' and made to leave the system 
from the moment that  it  escapes  
far downstream at the spatial boundary or if it reaches a well defined maximum energy. When a
number of particles escapes through the above mentioned boundaries, we replace them 
with new ones -in order to have again the same initial number of particles- but 
in such a way that as their weight is decreased, the number of particles remains almost
constant throughout the simulation.
Lorentz transformations are used between frames on shock passage. The results then are 
recorded just downstream, in the shock frame. For these simulations, we assume the shocks 
to be "parallel", either because that is the field configuration, or turbulence removes 
"reflection" at the interface.  We keep a compression ratio of 4 and 
consider both isotropic and pitch angle diffusion in the respective 
plasma rest frames.
The probability that a particle will  move a distance z 
along the field lines at pitch angle $\theta$ before a scattering
is given by the following expression:
\begin{equation}
Prob(z) \sim exp(-z/\lambda|cos\theta|)
\end{equation}
The new pitch angle $\theta'$, for pitch angle diffusion, is calculated 
by the simple trigonometric formula:
\begin{equation}
cos\theta'=cos\theta\sqrt{1-sin^2\delta\theta}+sin\delta\theta\sqrt{1-
cos^2\theta} cos\phi
\end{equation}
where $\phi$(0,2$\pi$) is the azimuth angle with respect to the original 
momentum direction. We note here that in the pitch angle diffusion case we keep 
the scattering within 
the 1/$\gamma_{up}$, where  $\gamma_{up}$ is the upstream gamma flow measured in the shock 
frame, in order to directly compare our results with theoretical investigations and predictions.}
\begin{center}

\large{\textit{\textbf {Results}}}
\end{center}

\begin{figure}
\begin{minipage}{170mm}
\epsfig{figure=plot.lg.g5,width=80mm}
\epsfig{figure=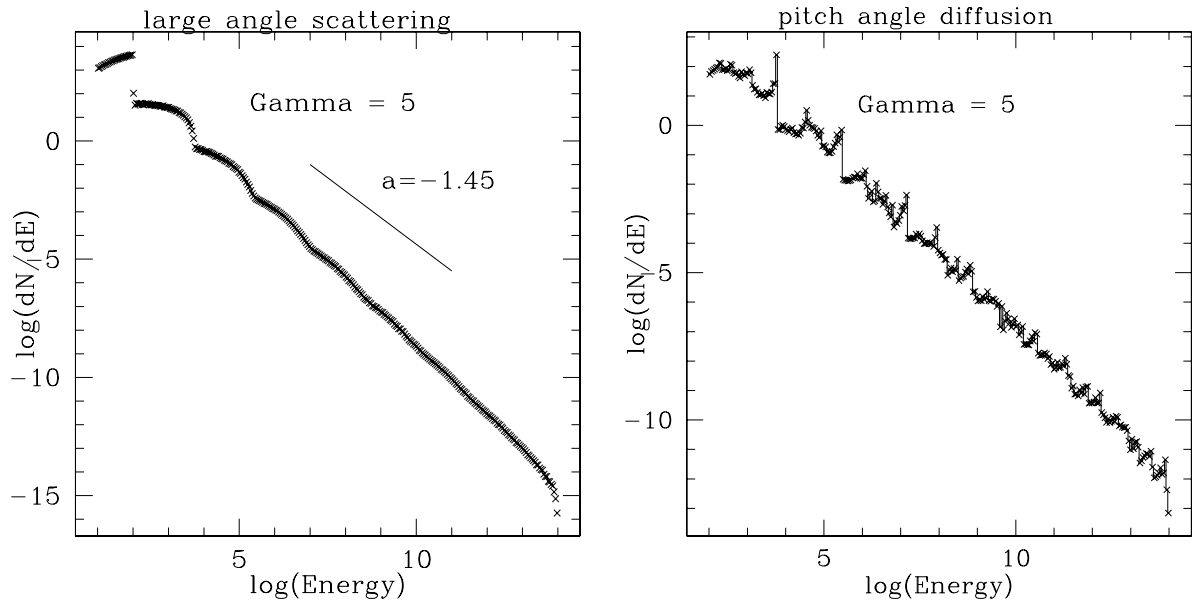,width=80mm}
\end{minipage}
\caption{Left plot--The \textit{smooth} spectral shape 
(for $\gamma$=5, large angle scattering) were the plateau 
structures slighlty appear near the top side of the spectrum. 
The plateau like stucture (due to consecutive cycling), and the spectral index is in 
consistency with similar computational work presented by Baring (1999).
Right plot--The spectral shape for the same gamma but for pitch angle diffusion.  
The same plateau-like spectrum structure hardly appears.}
\end{figure}

\begin{figure}
\begin{minipage}{170mm}
\epsfig{figure=plot.lg.g50,width=80mm}
\epsfig{figure=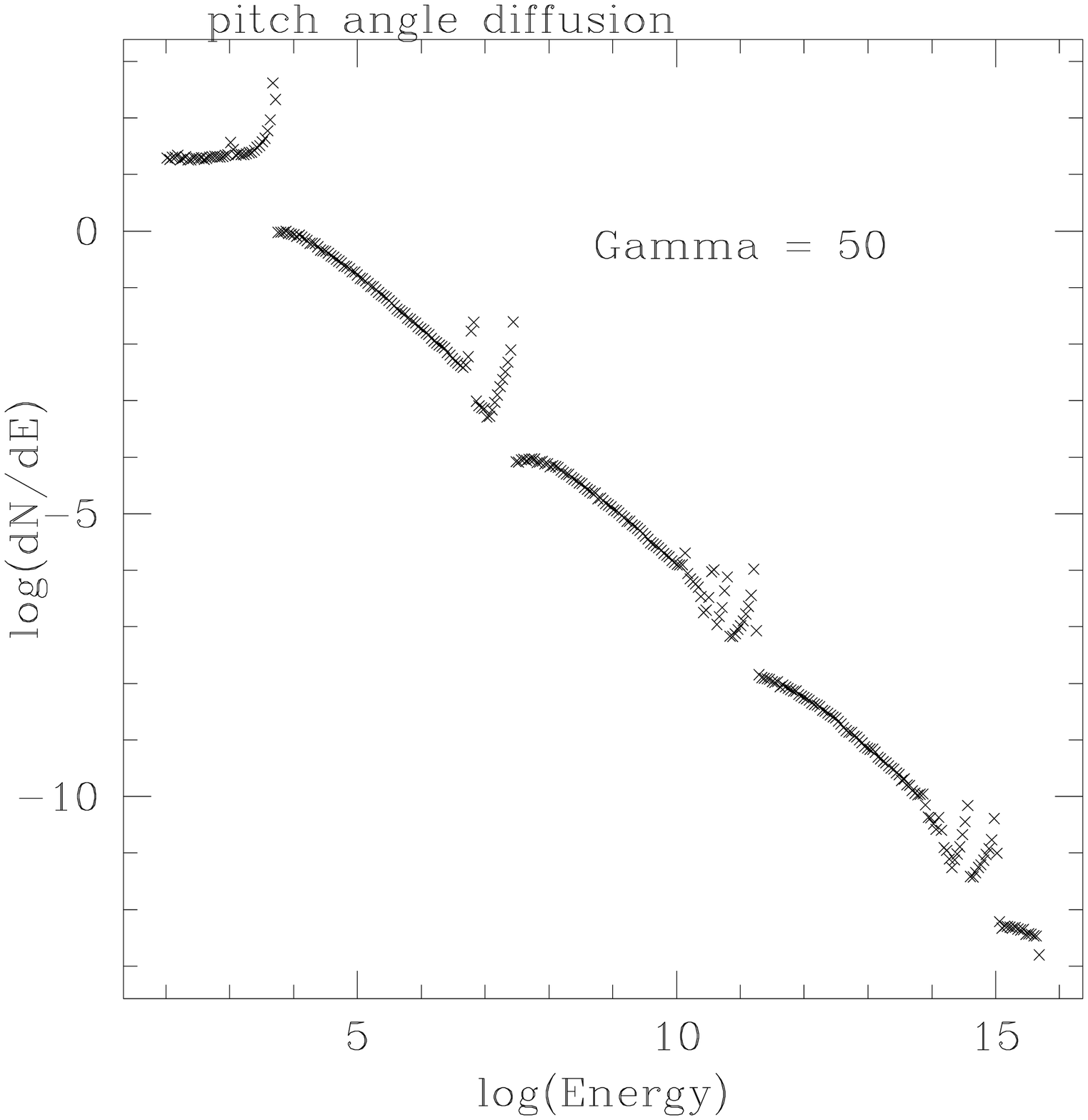,width=80mm}
\end{minipage}
\caption{Left plot--The spectral shape, for an upstream $\gamma$=50, 
for large angle scattering. The plateau like stucture appears due to
consecutive cycling of acceleration. Right plot- The equivalent spectral shape, for an
 upstream 
$\gamma$=50, for pitch angle diffusion. Dramatically enough, 
the effects of each discrete  acceleration cycle are even more evident with a 
plateau-structure steepening to be compared with large angle scattering.}
\end{figure}

\begin{figure}
\begin{minipage}{170mm}
\epsfig{figure=plot.lg.g200,width=80mm}
\epsfig{figure=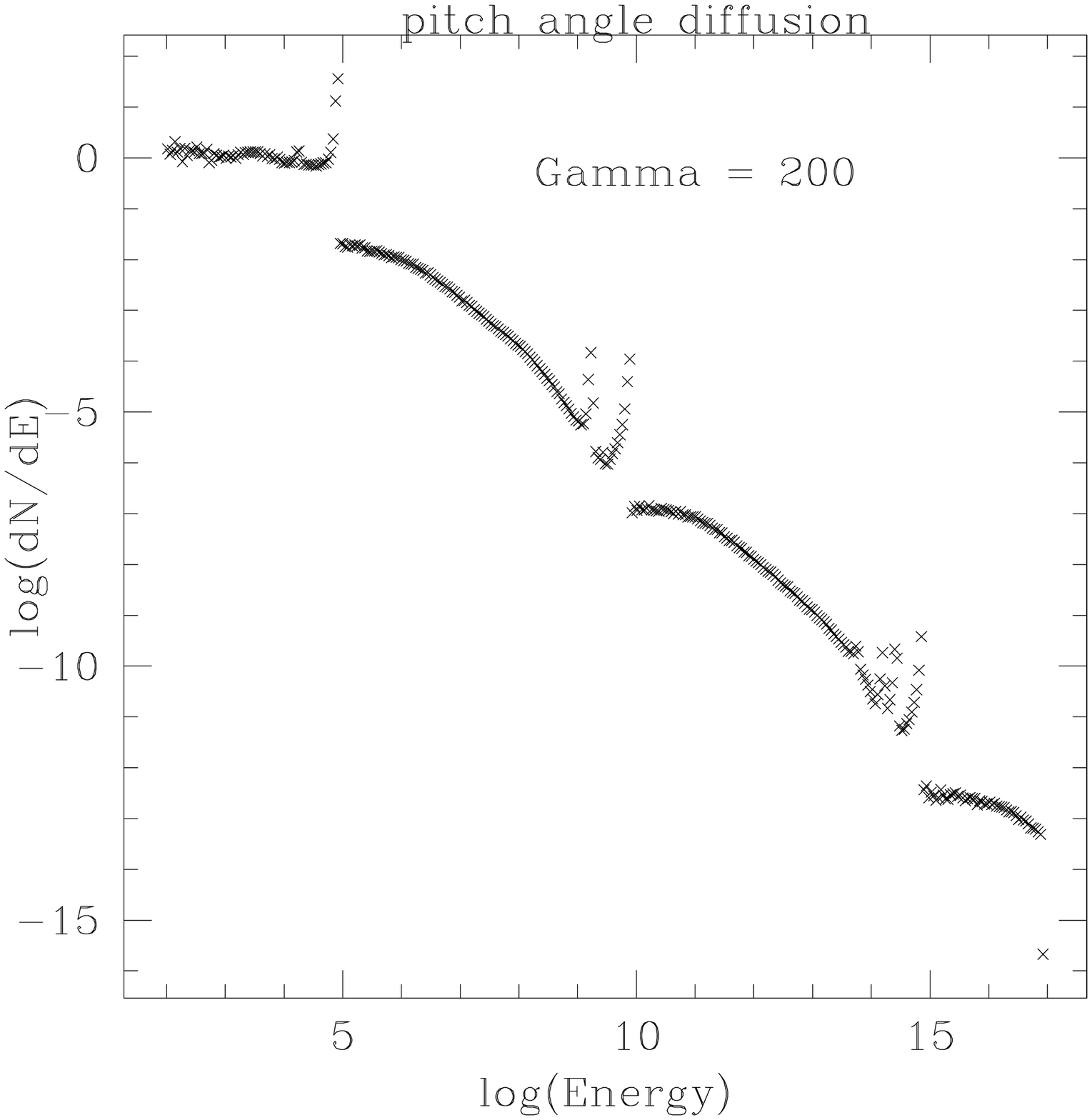,width=80mm}
\end{minipage}
\caption{Left plot--The spectral shape, for an upstream $\gamma$=200, 
for large angle scattering. The plateau like stucture appears due to
consecutive cycling of acceleration. Right plot-The equivalent spectral shape, for an upstream 
$\gamma$=200, for pitch angle diffusion. Here also,  
the effects of each discrete  acceleration cycle  are even more evident.}
\end{figure}

\begin{figure}
\begin{minipage}{170mm}
\epsfig{figure=plot.lg.g990,width=80mm}
\epsfig{figure=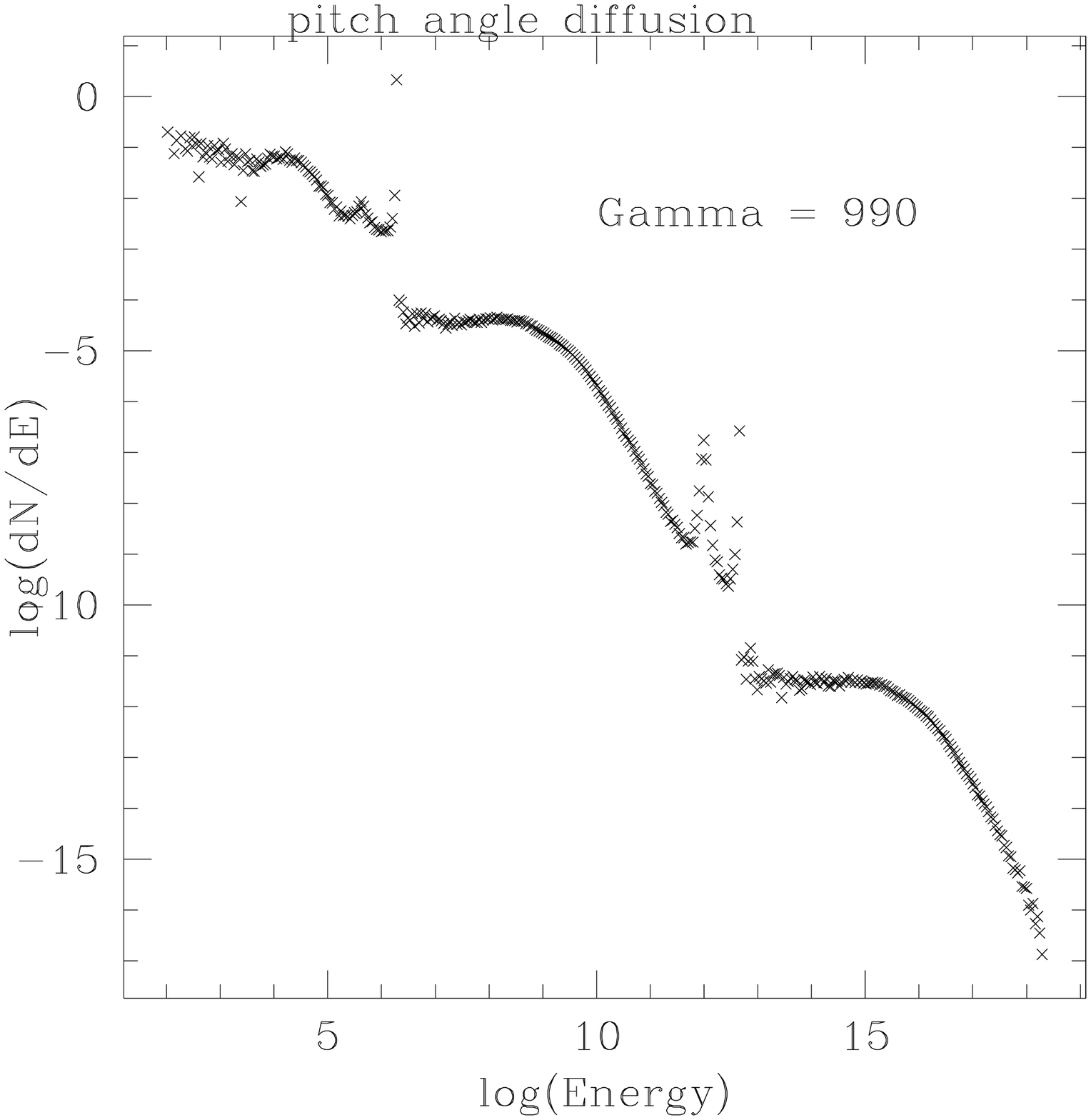,width=80mm}
\end{minipage}
\caption{Left plot--The spectral shape, for an upstream $\gamma$=990, 
for large angle scattering. Again, the plateau like stucture appears due to
consecutive cycling of acceleration. Right plot-- The equivalent spectral shape
for an upstream  $\gamma$=990, for pitch angle diffusion. Observe how the plateau parts of
spectrum steepen in comparison with the flat spectrum plateau parts in large angle scattering
 case.}
\end{figure}
{\large  The primary purpose of this work is to investigate the \textit{raw} spectral shape
and consequently the behaviour of particles being accelerated in 
ultra-relativistic shocks. These simulations are expected to be 
relevant within the context of a highly relativistic shock propagation
which could be found in GRBs. All results have been recorded in 
the downstream side, in the shock  frame. 
In the figures shown, there are results for isotropic scattering and 
pitch angle diffusion  for a various number of  upstream $\gamma$.
In figures 1, 2, 3 and 4 we can observe the very characteristic spectral shapes for 
large angle scattering and pitch angle diffusion. A distinctive feature is, the plateau-like 
structure for both large angle scattering and pitch angle diffusion mechanisms, 
at the left side and right side plots respectively, for Lorentz $\gamma$=5-990. 
It can be seen how the relativistic effects at each discrete acceleration cycle is 
evident, for both mechanisms.

The striking effect though is that the spectra do develop a smoother power-law shape 
\textit{in parts} which scales with the energy gain of the particles through the acceleration 
process. By observing closely the spectral shapes for large angle scattering and pitch angle
diffusion it is seen that in the case of pitch angle dissusion the plateau-like parts
of the spectrun -which correspond to the particle's shock cycling- is steeper than the plateau
structures of the large angle scattering mechanism. This implies that in pitch angle diffusion
case, there is an indication showing that the probability of convection downstream of the shock
drops much more with increasing energy than in the case of large angle scattering diffusion. This is a  
fact that has been shown theoretically in the relativistic regime, where for parallel shocks 
the large angle scattering gives flatter spectra than the pitch angle diffusion.   
So, one could infer that at relativistic flow velocities is difficult to assume
or take for granted that the spectral shape of the accelerated particles follows a 
\textit{smooth}
power-law shape although there is a characteristic monotonic increase in energy which is of course a signature of the 1st order Fermi acceleration. 
We can also understand the steepening behaviour partly  because  our calculations show that 
about $90\%$  of all particles in the simulation box are lost  downstream  each cycle.
We would like to note here that for the $\gamma=5$ for large angle scattering 
our spectral index has almost the same value (1.48) found in a similar work presented by Baring (1999) for $\gamma\sim$5.  
There is under way a computational approach and a more detailed study on 
parallel relativistic shocks and on highly relativistic 
oblique shocks which could be presented in a following paper.}

\begin{center}
{\large{\textit{\textbf {Conclusions}}}}
\end{center}

{\large A Monte Carlo numerical investigation has been reported in the 
 test particle limit of parallel diffusive shock acceleration. 
Very high gamma flow  astrophysical plasmas have been used, from $\gamma_{up}$ $\sim5$   
up to $\gamma_{up}$ $\sim1000$ which are relevant to AGNs but mostly to highly 
relativistic astrophysical supersonic 
plasma flows in GRBs. Very distinctive spectral shapes (plateau-like structures) have been 
demonstrated for both  large angle scattering and pitch angle diffusion and the 
spectral indexes  have been found to be in a good agreement with theoretial predictions, 
exhibiting a spectral flattening in the relativistic regime. These results reveal categorically a problem of the spectral shape of particles accelerated at ultra-relativistic shocks and provide fresh evidence regarding the behaviour of particles regarding the energy gain at every 
consecutive shock cycle concerning the large angle scattering and pitch angle diffusion 
mechanisms. This  crucial issue, needs further investigation as many other parameters 
need to be included  within the simulation codes in order for the models to be as realistic as 
possible.}
\newpage
\begin{center}
{\large{\textit{\textbf {References}}}}
\end{center}
{\small 
Baring, M.G, 1999, (ICRC Salt Lake City, Proceedings)\\
Kirk,J.G., \& Schneider, P., 1987a, ApJ 315, 425\\
Kirk, J.G, $\&$ Schneider, P., 1987, AJ 322, 256-265\\
Kirk,J.G., \& Webb, G.M., 1988, ApJ 331, 336\\
Lieu, R., \& Quenby, J.J., 1990, ApJ 350, 692\\
Meszaros,P., \& Rees, M.J., 1993, ApJ 405, 278\\
Peacock, J.A, 1981, \textit MNRAS 196, 135-152\\
Quenby, J.J., \&  Lieu, R.,1989,  Nature 342, 654\\

\end{document}